\documentclass[12pt]{iopart}

\usepackage{graphicx}

\newcommand\be{\begin{eqnarray}}
\newcommand\ee{\end{eqnarray}}
\newcommand\ba{\begin{array}}
\newcommand\ea{\end{array}}
\def\r{\rangle}
\def\l{\langle}
\def\T{{\rm Tr}}
\begin{document}
\title{Optimality of private quantum channels}
\author{Jan Bouda${}^{1,2}$ and Mario Ziman${}^{1,3}$}
\address{  ${}^{1}$
Faculty of Informatics, Masaryk University, Botanick\'a 68a, 602 00 Brno,
Czech Republic \\
${}^{2}$
Seibersdorf Research, Austria\\
${}^{3}$
Research Center for Quantum Information, Slovak Academy of Sciences,
Dubravsk\'a cesta 9, 845 11 Bratislava, Slovakia
}
\ead{xbouda1@fi.muni.cz,ziman@savba.sk}
\begin{abstract}
We addressed the question of optimality of private quantum channels.
We have shown that the Shannon entropy
of the classical key necessary to securely transfer the quantum information
is lower bounded by the entropy exchange of the private quantum channel
$\cal E$ and von Neumann entropy of the ciphertext state $\varrho^{(0)}$.
Based on these bounds we have shown that decomposition of
private quantum channels into orthogonal unitaries (if exists) is optimizing the entropy. For non-ancillary single qubit PQC we have derived the optimal entropy for arbitrary set of plaintexts. In
particular, we have shown that except when the (closure of the) set of 
plaintexts
contains all states, one bit key is sufficient. We characterized and analyzed
all the possible single qubit private quantum channels for arbitrary
set of plaintexts. For the set of plaintexts consisting of all qubit states
we have characterized all possible approximate private quantum channels
and we have derived the relation between the security parameter and the
corresponding minimal entropy.
\end{abstract}
\submitto{\JPA}
\section{Introduction}
Quantum cryptography \cite{gisin01,bouda04} (for a popular review see \cite{gottesman00}) is a rapidly developing branch of quantum information processing. The results of quantum cryptography include quantum key distribution \cite{bennett84,ekert91}, quantum secret sharing \cite{hillery99,cleve99}, quantum oblivious transfer \cite{bennett91,crepeau94} and other cryptographic protocols \cite{gruska99}. Quantum cryptography has two main goals: solutions to classical cryptographic primitives, and quantum cryptographic primitives.

The first goal is to design solutions of cryptographic primitives, which achieve a higher (provable) degree of security than their classical counterparts. The degree of security should be better than the security of any known classical solution, or it should be of the degree that is even not achievable by using classical information theory at all. Another alternative is to design a solution which is more efficient (according to time, space or communication complexity) than any classical solution of comparable security.

The second class of cryptosystems is motivated by the evolution of applications of quantum information processing, regardless whether their purpose is cryptographic, communication complexity based or algorithmic. These cryptosystems are designed to manipulate quantum information. As applications of quantum information processing start to challenge a number of their classical counterparts, the need to secure quantum communications in general is getting more urgent. Therefore, there is a large class of quantum primitives which should secure quantum communication in the same way as classical communication is secured. These primitives include encryption of quantum information using both classical \cite{ambainis00,boykin00,oppenheim03} and quantum key \cite{leung00}, authentication of quantum information \cite{barnum02}, secret sharing of quantum information \cite{cleve99,gottesman99}, quantum data hiding \cite{divincenzo02} and even commitment to a quantum bit \cite{bouda04}, oblivious transfer of quantum information \cite{bouda04} and others.

In this paper we concentrate on private quantum channels (PQC), i.e. on
schemes for perfect encryption of quantum information using 
a pre-distributed classical key, originally introduced in \cite{ambainis00}. 
Our aim is to analyze the optimal encryption of an arbitrary 
set of quantum states. In Section II we define the problem 
in general settings and investigate the elementary properties
of private quantum channels including their optimality.
Further, in Section III we focus on PQC for arbitrary set of qubit 
states. We will restrict ourselves to encryption schemes without ancillas.
The approximate private quantum channels (APQC) for a single qubit
are investigated in Section IV.
\section{Private quantum channels and optimality}
Consider a subset ${\cal P}\subset{\cal S}({\cal H})$
of quantum states. Its encryption is expressed by a
completely positive tracepreserving linear map
${\cal E}:{\cal B}({\cal H})\to{\cal B}({\cal H}\otimes{\cal H}_{\rm anc})$,
where ${\cal H}_{\rm anc}$ describes some ancillary system and 
${\cal B}({\cal H})$ stands for the set of bounded linear operators.
The encryption consists of two steps: i) addition of an ancilla 
in the state $\xi^{(j)}_{\rm anc}$ and ii) subsequent application of the 
unitary transformation $U_j$ with the probability $p_j$
satisfying the following identity 
\be
\label{pqc_def0}
{\cal E}[\varrho]=\sum_j p_j U_j
(\varrho\otimes\xi^{(j)}_{\rm anc})U_j^\dagger = \varrho^{(0)}
\ee
for all states $\varrho\in{\cal P}$. 
The triple $[{\cal P},{\cal E},\varrho^{(0)}]$
satisfying Eq.(\ref{pqc_def0}) defines a private quantum channel (PQC).
This definition establishes a communication quantum channel for which
an eavesdropper gains no information by intercepting the transmitted
messages. The indistinguishability of different convex decomposition 
of the same state ($\varrho^{(0)}$) guarantees the security of PQC.

The secure communication via PQC is successful only if the sender
and the receiver share the same classical key $(j_1,\dots,j_n)$ 
corresponding to the sequence of unitary operations $(U_{j_1},\dots,U_{j_n})$. 
The encryption as defined in Eq.\ref{pqc_def0} 
can be viewed as a noisy operation $\cal E$
caused by the environment interacting with the system and ancilla. 
Consequently, the decryption depends on our ability to inverse 
such noise. Therefore, only unitary decompositions of PQC channels 
are of interest \cite{werner01,nayak}. 
The decryption itself is similarly like encryption composed 
of two steps: i) application of a unitary operation according 
to shared classical key and ii) discarding of the ancilla.
The size of the classical key is quantified by its Shannon entropy 
($H=-\sum_j p_j\log_2 p_j$) and our aim is to address the question 
of the optimal encryption scheme
that enables us securely communicate an arbitrary set
of qubit plaintexts $\cal P$. The PQC with the smaller entropy
is better, because less amount of classical information must be
distributed prior to secure transmission of quantum states.
The authors in \cite{ambainis00} analyzed
two cases: i) encryption of all qubit states (optimal key $H=2$),
and ii) encryption of real superpositions (optimal key $H=1$),
i.e. the equatorial plane in the Bloch sphere picture.

Before we get to a specific case of single qubit PQC let us
introduce some important notions and properties of private
quantum channels. Due to linearity the PQC
transformation $\cal E$ encrypts not only the set $\cal P$, but
also arbitrary tracepreserving linear combination of its elements, i.e.
the whole set $\overline{\cal P}_{tp}
=\{ \varrho=\sum_k q_k \varrho_k:{\rm\ such\ that\ }
\varrho_k\in{\cal P}, \sum_k q_k=1, q_k {\rm\ is\ real}\}$.
The set of operations $U_j^\prime=V U_j$ ($V$ is unitary)
forms a PQC $[{\cal P},{\cal E}^\prime,\varrho^{(0)^\prime}]$ 
($\varrho^{(0)^\prime}=V\varrho^{(0)} V^\dagger$),
providing that the quadruple
$[{\cal P},{\cal E},\varrho^{(0)}]$
establishes a PQC for the plaintexts in $\cal P$. We note that ancilla
states $\xi_{\rm anc}^{(j)}$ and probabilities $p_j$ 
remain the same for both ${\cal E}^\prime,{\cal E}$.

In what follows we will restrict ourselves to PQC schemes
using a fixed ancilla state, i.e. 
$\xi_{\rm anc}^{(j)}=\varrho_{\rm anc}$ for all $j$. Thus for
PQC we have
\be
\label{pqc_def}
{\cal E}[\varrho]=\sum_j p_j U_j
(\varrho\otimes\varrho_{\rm anc})U_j^\dagger = \varrho^{(0)} \, .
\ee
Such restricted definition was used in the original work 
\cite{ambainis00} and it is sufficient for our purposes, because
in the next sections we will consider only PQC schemes without 
ancillas. The general case deserves a deeper investigation, but it is 
beyond the scope of this paper. Let us analyze the optimality for
 private quantum channels with a fixed state of the ancilla.

The entropy exchange function $S_{\rm ex}(\varrho,{\cal E})$
quantifies the amount of quantum information
lost in quantum environment due to interaction resulting in the
transformation $\cal E$ providing that the initial state of the
system is $\varrho$. In our case the state
$\varrho$ describes the system together with
the ancilla, because both of them are transmitted via the quantum channel
together. Due to Stinespring theorem each quantum channel
$\cal E$ can be expressed as a unitary transformation on a larger system, i.e.
${\cal E}[\varrho]=\T_{\rm env} G(\varrho\otimes |0\r\l 0|)G^\dagger
=\sum_j A_j\varrho A_j^\dagger$, where
$G$ is a unitary transformation describing the interaction with the
environment. The entropy exchange is defined as the von Neumann entropy of
the environment state after the interaction, i.e.
$S_{\rm ex}(\varrho,{\cal E})=S(\omega_{\rm env})$
with $\omega_{\rm env}=\T_{\rm system}[G(\varrho\otimes |0\r\l 0|)G^\dagger]=
\sum_{jk} \T[A_j\varrho A_k^\dagger] |j\r\l k|$. This quantity does not depend
on particular Kraus representation, because different Kraus representations
of the same quantum channel are related by unitary transformation.
Since PQCs are always random unitary channels, it follows that
$\omega_{\rm env}
=\sum_{jk} \sqrt{p_j p_k} \T[U_j\varrho U_k^\dagger] |j\r\l k|$.
From the definition of von Neumann entropy \cite{perez} as the minimum
of the Shannon entropy over all projective measurements
the following inequality holds
$S(\omega_{\rm env})\le S({\rm diag}_{\cal B}[\omega_{\rm env}])$ 
for arbitrary state
$\omega_{\rm env}$. Thus, the equality is achieved if the 
basis $\cal B$ coincides with the eigenbasis of the density 
operator $\omega_{\rm env}$. The operation
${\rm diag}_{\cal B}$ cancels all off-diagonal terms in the description
of the density matrix in the basis $\cal B$. In our case we have
$S({\rm diag}[\omega_{\rm env}])=H(\{p_k\})$, i.e. entropy of the shared key
equals to entropy of the diagonal elements of the state $\omega_{\rm env}$.
Hence, we obtain the following lower bound on the key entropy
\be
\label{key_bound1}
H(\{p_k\})\ge \max_\varrho S_{\rm ex}(\varrho,{\cal E})\, .
\ee
The left side of this inequality does depend on the particular
convex decomposition of the quantum channel $\cal E$, but the right hand
side is independent of the particular realization of $\cal E$.
Therefore, the smallest possible entropy of the key
$H$ is given by the maximum of the entropy exchange.
We have seen that this inequality is saturated
only if the environment state is diagonal, i.e. $\T [U_j\varrho U_k^\dagger]=0$
for $j\ne k$. Choosing $\varrho\sim I$ we obtain the orthogonality
condition for unitary transformations $U_j$, i.e. $\T[U_j U_k^\dagger]=0$
for $j\ne k$. Thus, convex decompositions into orthogonal unitary
transformations saturates the above inequality. The right side is independent
of the Kraus representation (decomposition) of ${\cal E}$ and, moreover,
for random unitary channels the maximum of $S_{\rm ex}$ is achieved
for total mixture, because the diagonal elements are independent
of the state $\varrho$, i.e.
${\rm diag} [\omega_{\rm env}]={\rm diag}
[\sum_{jk}\sqrt{p_j p_k}\T(U_j\varrho U_k^\dagger)|j\r\l k|]=
\sum_j p_j |j\r\l j|$ for all states $\varrho$.
As a result we have obtained that the optimal realization of PQC
minimizing the entropy of classical key is achieved for
the encryption with mutually orthogonal unitary
transformations. Open question is whether such orthogonal decomposition
exists for all random unitary channels, or at least for all PQCs.

For a mixture of pure states $\varrho=\sum_j p_j |\psi_j\r\l \psi_j|$
the following inequality holds $S(\varrho)\le H(\{p_j\})$. Consider a pure
state $|\psi\r\in\overline{{\cal P}}_{\rm tp}$. Encryption operation
$\cal E$ results in a mixture $\varrho^{(0)}=\sum_j p_j |\psi_j\r\l\psi_j|$
with $|\psi_j\r= U_j|\psi\r$, i.e. we can use the entropy of the state
$\varrho^{(0)}$ to bound the entropy of the key from below
\be
H(\{p_j\})\ge S(\varrho^{(0)}) \, .
\ee
The previous lower bound in Eq.(\ref{key_bound1}) determines the optimal
value of the classical key entropy for a given PQC $\cal E$, but this bound
enables us to limit the key entropy of PQC based on the state
$\varrho^{(0)}$. This inequality suggests that the smaller the entropy of
$\varrho^{(0)}$, the more optimal PQC could exist.
For a given set of plaintexts this means that the most optimal PQC
should be the one with the purest possible state $\varrho^{(0)}$.
However, this bound is not achievable in general. For instance, the
encryption of all single qubit states requires $H=2$, but the entropy
of the maximally mixed single qubit state is
$S(\varrho^{(0)}=\frac{1}{2}I)=1$.

\section{Single qubit private quantum channels}
In the previous section we have
shown that the optimal realization of a fixed PQC consists of orthogonal
unitary transformations. However, the more general question is the optimal
PQC for a given set of plaintexts $\cal P$. The first question is to specify
the states $\varrho^{(0)}$ achievable by PQC. As we have argued at the
end of the previous section the purer the state $\varrho^{(0)}$, the smaller
could be the entropy of the classical key. Using the fact that arbitrary
tracepreserving linear map cannot increase the trace distance between two
states ($D(\varrho,\sigma)=\T|\varrho-\sigma|$) we obtain
\be
\delta=\min_{\varrho\in\overline{\cal P}_{\rm tp}}
D\left(\varrho\otimes\varrho_{\rm anc},\frac{1}{N}I\right)
\ge D\left(\varrho^{(0)},\frac{1}{N}I\right)\, ,
\ee
where $N={\rm dim}({\cal H}\otimes {\cal H}_{\rm anc})$ is the dimension of
the system together with the ancilla. This inequality restricts the
possible states $\varrho^{(0)}$ to the $\delta$
vicinity around the total mixture $\frac{1}{N}I$, but the achievability
of all such states must be proved, see below.

\subsection{Single qubit ancilla-free PQCs}
The PQC channel is a special random unitary channel, hence it is unital
(preserves the total mixture $\frac{1}{N}I$). This feature makes it easy to
analyze all PQC for a single qubit, i.e. for two-dimensional quantum system,
if we restrict ourselves to PQC without ancillas (ancilla-free PQC). Under
such condition the PQCs are just single qubit unital channels that coincide
with random unitary channels, i.e. each single qubit unital channel
can be expressed as a convex combination of unitary transformations
\cite{ruskai}. In what follows we will use the Bloch sphere representation
of qubit states, i.e. as a three-dimensional real vectors $\vec{r}$
specifying the state $\varrho=\frac{1}{2}(I+\vec{r}\cdot\vec{\sigma})$, where
$\vec{\sigma}=(\sigma_x,\sigma_y,\sigma_z)$ is the vector of Pauli operators.
The unital quantum channels $\cal E$ correspond to Bloch vector transformations
$\vec{r}\to\vec{r}^\prime=T\vec{r}$, where $T$ is a 3x3 real matrix with
coefficients $T_{jk}=\T (\sigma_j{\cal E}[\sigma_k])$. Each unital
quantum channel can be written via two unitary transformations
$U,V$ and a specific quantum
channel $\Phi_{\cal E}$ in the following way ${\cal E}[\varrho]
=U\Phi_{\cal E}[V^\dagger\varrho V]U^\dagger$. The operation $\Phi_{\cal E}$
is just the convex combination of mutually orthogonal unitary transformations
(Pauli operators $I,\sigma_x,\sigma_y,\sigma_z$), i.e.
$\Phi_{\cal E}[\varrho]=\sum_j p_j \sigma_j\varrho\sigma_j$.
In Bloch sphere representation this corresponds to the product
of three matrices (singular value decomposition)
$T=R_U D R_V$, where $R_U,R_V$ are the corresponding rotations of the Bloch
sphere around its origin and $D={\rm diag}\{\lambda_1,\lambda_2,\lambda_3\}$
is a diagonal matrix with $\lambda_j$ being singular values of the matrix $T$
(for details see \cite{ruskai,bouda05}). They are related to
the probabilities $p_j$ via the following identities
\be
\label{prob_lambda}
\ba{rcl}
p_x&=&\frac{1}{4}(1+\lambda_1-\lambda_2-\lambda_3)\\
p_y&=&\frac{1}{4}(1-\lambda_1+\lambda_2-\lambda_3)\\
p_z&=&\frac{1}{4}(1-\lambda_1-\lambda_2+\lambda_3)\\
p_0&=&1-p_x-p_y-p_z \, .
\ea
\ee
To guarantee the positivity the parameters $\lambda_1,\lambda_2,\lambda_3$
are constrained by the inequalities $p_j\ge 0$.

In the previous section
we have shown that orthogonal decompositions of a given channel $\cal E$
are the optimal ones (in the sense of the key entropy). Moreover, the entropy
of the classical keys are the same for quantum channels $\cal E$ and
$\Phi_{\cal E}$ (they are unitarily equivalent). Therefore it is
sufficient to analyze only the so-called
Pauli channels $\Phi_{\cal E}$, for which the optimal
realization is clearly a convex combination of Pauli
unitary rotations. Let us start with the specification of possible
sets of plaintexts $\overline{\cal P}_{\rm tp}$. The smallest possible
set consists of a single plaintext (${\cal P}^1=\{\varrho\}$), but in
this case the situation is trivial, because there is nothing to
hide. We remind us that the set of plaintexts is publicly known.
The largest possible set contains all qubit states, i.e.
${\cal P}^4={\cal S}({\cal H})$. We can meet with such case whenever
the set ${\cal P}$ contains four (not only three!)
mutually independent quantum states (Bloch vectors). The mutual
independence means that one of them cannot be written as some
tracepreserving linear combination of the others. The tracepreserving
linear combinations form a set covering the whole Bloch sphere,
i.e. $\overline{\cal P}_{\rm tp}={\cal S}({\cal H})$.
For this maximal possible set of plaintexts the optimal private quantum channel
is represented by the completely depolarizing channel mapping all
states into the total mixture, i.e. $\Phi_{\cal E}[\varrho]=
\frac{1}{4}(\varrho+\sigma_x\varrho\sigma_x+\sigma_y\varrho\sigma_y
+\sigma_z\varrho\sigma_z)=\frac{1}{2}I$ for all $\varrho$. Thus
a classical key of the length of two bits is
necessary for the encryption of the whole Bloch sphere, i.e. $H=2$
\cite{ambainis00}.

Our aim is to investigate the minimal
length of the classical key for other possible sets of plaintexts $\cal P$.
In principle, there are only two remaining options:
the set $\overline{\cal P}_{tp}$
is generated either by two states (${\cal P}_2=\{\varrho_1,\varrho_2\}$),
or by three states $({\cal P}_3=\{\varrho_1,\varrho_2,\varrho_3\})$.
The goal is to find the dependence of the key entropy on particular
properties of these generating states. In the Bloch sphere picture
the sets  $\overline{\cal P}^2_{\rm tp}$ and $\overline{\cal P}^3_{\rm tp}$
can be illustrated as lines and planes, respectively,
intersecting the Bloch sphere (for details see \cite{bouda05}).
It is known \cite{ambainis00}
that for the so-called real qubits, i.e. real superpositions of two
orthogonal pure states, only single bit is
sufficient to establish a private quantum channel.
Such states form a particular set of plaintexts consisting of
all equatorial states of the Bloch sphere. Using this result
we can conclude that
there exists a private quantum channel for arbitrary set
$\overline{\cal P}^2_{\rm tp}$ with the entropy $H=1$. This
can be seen directly from the Bloch sphere representation, because
real superpositions form a circle containing the center of the Bloch sphere,
but each line associated with $\overline{\cal P}^2_{\rm tp}$ belongs to
some plane containing the total mixture. Therefore, the real
qubit encryption PQC scheme works for all lines belonging to the corresponding
real qubit plane. However, it is not known whether for a specific set
of plaintexts (not containing the total mixture) we cannot do better
and establish a PQC with $H<1$.

\subsection{All ancilla-free PQCs}
In what follows we will address the following question: which single qubit
unital maps constitute a PQC? Except the trivial case (${\cal P}^{1}$),
the set $\overline{\cal P}_{tp}$ contains at least two pure
states $|\psi_1\r,|\psi_2\r\in\overline{\cal P}_{tp}$. The PQC maps these
two states into the state $\varrho^{(0)}$. Using the Bloch sphere
representation this means that $\vec{r}_1\to\vec{r}_1^\prime=\vec{s}$
and $\vec{r}_2\to\vec{r}_2^\prime=\vec{s}$, where $\vec{r}_1,\vec{r}_2$
correspond to pure states, respectively, and $\vec{s}$ is associated with
$\varrho^{(0)}$. Using the explicit form of Pauli channels
($\Phi_{\cal E}\leftrightarrow D={\rm diag}\{\lambda_x,\lambda_y,\lambda_z\}$)
the identity $\vec{r}_1^\prime-\vec{r}_2^\prime=\vec{0}$ results in the
system of equations $\lambda_j(r_{1j}-r_{2j})=0$ for all components
$j=x,y,z$. This equalities holds only if $\lambda_j=0$, or
$r_{1j}=r_{2j}$. If none of the $\lambda$s vanishes
($\lambda_x\lambda_y\lambda_z\ne 0$), then necessarily
$\vec{r}_1=\vec{r}_2$, i.e. the states are identical. Therefore at least
one of the $\lambda_j$ must vanish. Choose $\lambda_z=0$. The complete
positivity constraint \cite{ruskai} restricts the possible values of
$\lambda_x,\lambda_y$ so that the inequality $|\lambda_x\pm\lambda_y|\le 1$
specifies all possible (non-ancillary) single qubit private quantum channels,
i.e. the general single qubit PQC is up to
a unitary rotation represented in its optimal form as follows:
\be
{\cal E}[\varrho]=\frac{1}{4}( (1+b)\varrho+(1+a)\sigma_x\varrho\sigma_x+
(1-a)\sigma_y\varrho\sigma_y+(1-b)\sigma_z\varrho\sigma_z)
\ee
such that $a=\lambda_x-\lambda_y, b=\lambda_x+\lambda_y$ and complete
positivity constraints $|a|\le 1, |b|\le 1$. The (optimal) entropy of the
classical key necessary for establishing the general PQC channel equals
\be
H({\cal E})&=&2-\frac{1}{4}\left[ h(a)+h(b)\right] \, ,
\ee
where $h(x)=(1+x)\log(1+x)+(1-x)\log(1-x)$.

\subsection{Two linearly independent states}
Consider a set of plaintexts
${\cal P}_{xy}=\{\varrho_z=\frac{1}{2}(I+x\sigma_x+y\sigma_y+z\sigma_z)\}$
with $x,y$ fixed. Each PQC given by $D={\rm diag}(\lambda_x,\lambda_y,0)$
enables us to transmit these sets securely.
In the Bloch sphere representation
these sets form lines parallel to the line connecting
the poles of the Bloch sphere. Without loss of generality we can assume
that sets $\overline{\cal P}_{xy}$ are the most general sets of type
$\overline{\cal P}^{2}_{\rm tp}$. Indeed, arbitrary set
$\overline{\cal P}^{2}_{\rm tp}$ is just a unitarily rotated set
$\overline{{\cal P}_{xy}}=\{\varrho_z=\frac{1}{2}(I+x\sigma_x+y\sigma_y+z\sigma_z)\}$
for some values $x,y$.  In particular, given a set ${\cal P}^2$
as a segment of the line $l$ crossing the Bloch sphere,
it is always possible to choose the coordinate system in the following way:
the $x$ axis is given by the center of the Bloch sphere ($\frac{1}{2}I$)
and the middle point of the segment of the line $l$ (most mixed state in
$\overline{P}^2_{\rm tp}$), the $y$ axis is perpendicular to the plane
given by the whole line $l$ and the total
mixture. This choice of the new coordinates
corresponds to a unitary rotation of the Pauli
operators $\sigma_j\to S_j=U\sigma_j U^\dagger$. In this basis the line
is given by the states
$\varrho_{z^\prime}=\frac{1}{2}(I+x^\prime S_x+z^\prime S_z)$
for some fixed $x^\prime$. For instance, the states in ${\cal P}_{xy}$
can be transformed into this form by a suitable rotation around the $z$ axis.
Therefore, the analysis of
$\overline{\cal P}_{\rm tp}^2$ reduces to the analysis of this type
of states. Using the expression for the distance between an arbitrary state
$\varrho\leftrightarrow\vec{r}$ and the total mixture
$D(\varrho,\frac{1}{2}I)=|\vec{r}|$ it follows that the closest
state from the general set of plaintexts $\overline{\cal P}_{\rm tp}$
is always the one associated with the shortest Bloch vector. For
the states of the form $\varrho_{z^\prime}$ the minimum is achieved
for $z^\prime=0$, i.e. for the state
$\varrho_{\min}=\frac{1}{2}(I+x^\prime S_x)$,
for which $D(\varrho_{\min},\frac{1}{2}I)=\delta=|x^\prime|$. The solution
for real qubits guarantees the existence of PQC with $H=1$ and
$\varrho^{(0)}=\frac{1}{2}I$ (i.e. $\delta=0$), but PQCs for other states
$\varrho^{(0)}$ are possible as well. Nevertheless, the formula for
the  entropy for general PQC channel guarantees that $H$ cannot be
smaller than one, i.e. $H\ge 1$. Although we cannot improve
the entropy rate for $\overline{\cal P}^2_{\rm tp}$
we are still curious about the possibility
$\varrho^{(0)}=\varrho_{\min}$. A direct calculation gives us that
the encryption $(\frac{1}{2},I)(\frac{1}{2},S_x)$ establishes a PQC
with the desired property, i.e.
\be
\varrho^{(0)}=\frac{1}{2}(\varrho_{z^\prime}
+S_x\varrho_{z^\prime}S_x)=\frac{1}{2}(1+x^\prime S_x)=\varrho_{\min}
\ee
with the entropy of the key $H=1$. In fact, arbitrary state within the
sphere $|\vec{r}|\le\delta=|x^\prime|$ is achievable with the entropy
$H=1$.

\subsection{Three linearly independent states}
The sets
${\cal P}_x=\{\varrho_{yz}=\frac{1}{2}(I+x\sigma_x+y\sigma_y+z\sigma_z)\}$
for a fixed $x$ and arbitrary $y,z$ represent up to unitary rotations
the most general sets of plaintexts of the type
$\overline{\cal P}^3_{\rm tp}$, i.e.
planes in the Bloch sphere representation perpendicular to $x$ axis.
In this case not all PQC
(up to unitary rotation) are suitable, but only those, for
which $\lambda_y=\lambda_z=0$ and $\lambda_x\ne 0$. This class
of PQCs is more powerful, because it encrypts not only the
sets $\overline{\cal P}^2_{\rm tp}$, but also the sets
$\overline{\cal P}^3_{\rm tp}$. These channels correspond to
so-called phase damping channels, i.e. they describe the
most general pure decoherence processes \cite{ziman05}.
Using a suitably rotated PQC
of this form ($D={\rm diag}\{\lambda_x,0,0\}$) we can encrypt
any possible set $\overline{\cal P}_{\rm tp}^2$ and
$\overline{\cal P}_{\rm tp}^3$ with the entropy
\be
\label{entropy3}
H=2-\frac{1}{2}[(1+\lambda_x)\log(1+\lambda_x)+(1-\lambda_x)\log(1-\lambda_x)]\, .
\ee
It follows that the smallest possible value of the entropy is the same
for both types of sets and equals $H=1$, i.e. except the plaintexts
containing the whole set of states $\overline{\cal P}_{\rm tp}
={\cal S}({\cal H})$, a single bit classical key is sufficient to establish
a private quantum channel transmitting all plaintexts
$\varrho\in\overline{\cal P}_{\rm tp}$ for arbitrary set $\cal P$, for which
$\overline{\cal P}_{\rm tp}\ne {\cal S}({\cal H})$.

\begin{figure}
\begin{center}
\includegraphics[width=10cm]{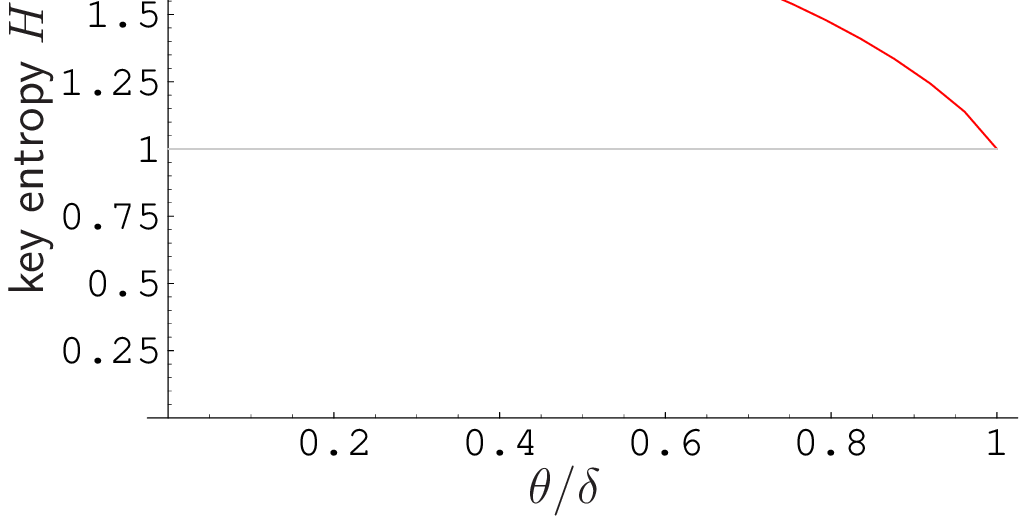}
\caption{Optimal entropy as a function of the distance
$\theta=D(\varrho^{(0)},\frac{1}{2}I)$ for arbitrary set of plaintexts
characterized by parameter $\delta=
\min_{\varrho\in\overline{\cal P}_{\rm tp}}D(\varrho,\frac{1}{2}I)$.
The upper line describes the dependence for sets of plaintexts containing
three independent states (planes) and the lower (constant) line describes
the situation for plaintexts containing only two independent states (lines).
}
\end{center}
\end{figure}

\subsection{Optimality}
The optimal value is achieved for $\lambda_x=1$, i.e. the corresponding
private quantum channels are unitarily equivalent to
\be
\label{pqc_xopt}
\Phi_{\cal E}^{\rm opt}[\varrho]
=\frac{1}{2}(\varrho+\sigma_x\varrho\sigma_x)\, .
\ee
We have analyzed the achievability of states $\varrho^{(0)}\ne\frac{1}{2}I$
for sets $\overline{\cal P}^2_{\rm tp}$ and the question is whether
the situation is similar as it was in the case of sets
$\overline{\cal P}^3_{\rm tp}$.
In particular, $\varrho_{\min}=\frac{1}{2}(I+x\sigma_x)$
is the state with the smallest length of the Bloch vector among
all states in ${\cal P}_x$. Is it possible to design a PQC
such that $\varrho^{(0)}=\varrho_{\min}$? The answer is simple, because
the PQC given in Eq.(\ref{pqc_xopt}) satisfies this property, i.e.
$\Phi_{\cal E}^{\rm opt}[{\cal P}_x]=\frac{1}{2}(I+x\sigma_x)$. Since all
other sets $\overline{\cal P}^3_{\rm tp}$ are just unitarily rotated
sets ${\cal P}_x$, it follows that the states $\varrho_{\min}$
are achievable in general. For a general PQC
$[\overline{\cal P}^3_{\rm tp},D={\rm diag}\{\lambda_x,0,0\},\varrho^{(0)}]$
the allowed states $\varrho^{(0)}$ are inside the sphere determined by the
condition $|\vec{r}|\le\delta$. In particular,
$D(\varrho^{(0)},\frac{1}{2}I)=|\lambda_x x|=|ax|\le \delta$ and
the entropy $H$ is given by formula in Eq.(\ref{entropy3}), i.e. it increases
as the distance $D(\varrho^{(0)},\frac{1}{2}I)$ is decreasing. Denote by
$\theta$ the distance $D(\varrho^{(0)},\frac{1}{2}I)$. Then for a given
value of $\theta$ the corresponding PQC transformation is
$D={\rm diag}\{\theta/\delta,0,0\}$.

As a result we have derived that the optimal entropy of the classical key
for arbitrary (two or three dimensional) set of plaintexts equals $H=1$. Moreover, we have found the
dependence of the optimal entropy
on the distance $\theta$ between the state $\varrho^{(0)}$
and the total mixture (see also Fig.1)
\be
H(\overline{\cal P}^2_{\rm tp},\theta)&=& 1\\
H(\overline{\cal P}^3_{\rm tp},\theta)&=& 2-\frac{1}{2}
[(1+\theta/\delta)\log(1+\theta/\delta)+(1-\theta/\delta)\log(1-\theta/\delta)] \, .
\ee
\section{Approximate private quantum channels}
Approximate private quantum channels (APQC) generalize the ideal version in
the following sense. The quadruple
$[{\cal P},{\cal E},\varrho_{\rm anc},\epsilon]$ constitutes
an $\epsilon$-private quantum channel ($\epsilon$-PQC), if
\be
D({\cal E}[\varrho_1\otimes\varrho_{\rm anc}],
{\cal E}[\varrho_2\otimes\varrho_{\rm anc}])\le\epsilon
\ee
for all $\varrho_1,\varrho_2\in{\cal P}$. As before, the encryption
operation $\cal E$ consists of a mixture of unitary transformations
$U_j$ applied with probabilities $p_j$. Similarly, the decryption
operation is given by application of the inverse operations $U_j^\dagger$
according to a shared classical key represented by the sequence of
unitaries $U_{j_1},\dots,U_{j_n}$. In such generalization of PQC
the transmission is still perfect and $\epsilon$ quantifies
the security of the protocol, i.e. the distinguishability of the
transferred states.

We are not going to discuss the problem of optimality for such generalization
in its full generality, but we will pay attention to encryption of the
qubit states without using any additional ancillas. The set of all
approximate private quantum channels is a specific subset of all
random unitary channels determined by the value of $\epsilon$.
As we have mentioned, arbitrary single qubit unital channel can be written as a
convex combination of unitary transformations and upto
a unitary transformations the general unital qubit channel
$\cal E$ is specified by three parameters $\lambda_x,\lambda_y,\lambda_z$
related to probabilities $p_0,p_x,p_y,p_z$ via the Eq.(\ref{prob_lambda}).
The entropy achieves its optimal value (minimum) for the orthogonal
unitary decomposotion of $\cal E$, hence it equals the entropy
of this probability distribution, $H=-\sum_j p_j\log p_j$.

It follows that each unital qubit channel establishes
an $\epsilon$-PQC, but we still need to specify the particular
value $\epsilon$ and then analyze the optimal entropy
as a function of the degree of privacy for different sets of plaintexts
$\cal P$. Let us analyze the case when the set of plaintexts $\cal P$
consists of all quantum states, i.e. $\cal P$ equals
Bloch sphere. For a given PQC $\cal E$ (i.e. arbitrary unital channel)
we have
\be
\nonumber
\epsilon &=& \max_{\varrho_1,\varrho_2\in{\cal S}(\cal H)}
D({\cal E}[\varrho_1],{\cal E}[\varrho_2])=
\max_{\vec{r}_1,\vec{r}_2} \sqrt{\sum_j
\lambda_j^2|r_{1j}^\prime-r_{2j}^\prime|^2}\\
&=&2\max\{|\lambda_x|,
|\lambda_y|,|\lambda_z|\}\equiv 2\lambda_{\max} \, .
\ee
Without loss of generality we can assume that
$|\lambda_x|\le|\lambda_y|\le|\lambda_z|=\epsilon/2$. Our aim is to analyze
and relate the functions $\epsilon=\epsilon(\lambda_x,\lambda_y,\lambda_z)$
and $H=H(\lambda_x,\lambda_y,\lambda_z)$. In particular we are interested
in two questions: i) given $\epsilon$ what is the optimal approximative
PQC, i.e. with the minimal entropy $H$, and ii) given the entropy $H$ what is
the most perfect PQC, i.e. with the smallest $\epsilon$.

Consider $\lambda_z$ is fixed and $|\lambda_x|\le|\lambda_y|\le|\lambda_z|$,
i.e. the security parameter $\epsilon=2|\lambda_z|$ is fixed.
For unital single qubit channels the conditions of complete positivity
(see Eq.(\ref{prob_lambda})) read
\be
\nonumber 1-\lambda_z&\ge& \pm (\lambda_x-\lambda_y) \\
 1+\lambda_z&\ge& \pm(\lambda_x+\lambda_y) \, .
\label{ineq}
\ee
The values $\lambda_x,\lambda_y,\lambda_z$ satisfying these conditions
form a tetrahedron with vertices associated with the
orthogonal unitary transformations
$I,\sigma_x,\sigma_y,\sigma_z$. Since these vertices are
unitarily related, the whole tetrahedron can be divided into four
unitarily equivalent parts containing channels with the same
values of entropy. It follows
that it is sufficient to analyze only two regions (forming a particular
single part): with strictly positive values
($\lambda_x,\lambda_y,\lambda_z\ge 0$)
and with strictly negative values ($\lambda_x,\lambda_y,\lambda_z\le 0$).

Intuitively, the geometric picture
suggests that the most optimal APQC should shrink the Bloch
sphere symmetrically,
i.e. $|\lambda_x|=|\lambda_y|=|\lambda_z|=\lambda$. However, not for all
positive, or negative combinations of $\lambda_x,\lambda_y,\lambda_z$
such transformation is associated with some completely positive map, i.e.
the probabilities $p_j$ in Eq.(\ref{prob_lambda}) are not positive.
Let us assume that $\lambda_x=\lambda_y=\lambda_z=\lambda$ then
\be
H_I=2-\frac{1}{4}[(1+3\lambda)\log(1+3\lambda)+3(1-\lambda)\log(1-\lambda)]\, .
\label{ent_depol}
\ee
However, the transformation is physical (and entropy makes sense) only if
$-\frac{1}{3}\le\lambda\le 1$. Using the relation $\epsilon = 2|\lambda|$
we obtain the dependence of the entropy on the approximation
(security) parameter $\epsilon$ (see Fig.2) for this class of channels.
As we can see from Fig.2 for smaller values of the security parameter
$\epsilon$ the channels given by negative values of $\lambda$ are more optimal.
Indeed, one can test numerically that these points are optimal among all possible private quantum channels for
$\epsilon\in[0,2/3]$. Let us note that  the point $\lambda=-1/3$ corresponds
to the best physical approximation of the universal NOT operation
(given by unphysical values $\lambda_x=\lambda_y=\lambda_z=-1$). At this point
the entropy equals $H=1.585$. In the interval $2/3\le\epsilon\le 0.958$
the optimal entropy is constant because the optimal depolarizing channel
is still the best approximation of the universal NOT ($\lambda=-1/3$).
After that interval the entropy decreases again until its minimal
value $H=0$ is achieved (for $\epsilon=2$).

\begin{figure}
\begin{center}
\includegraphics[width=10cm]{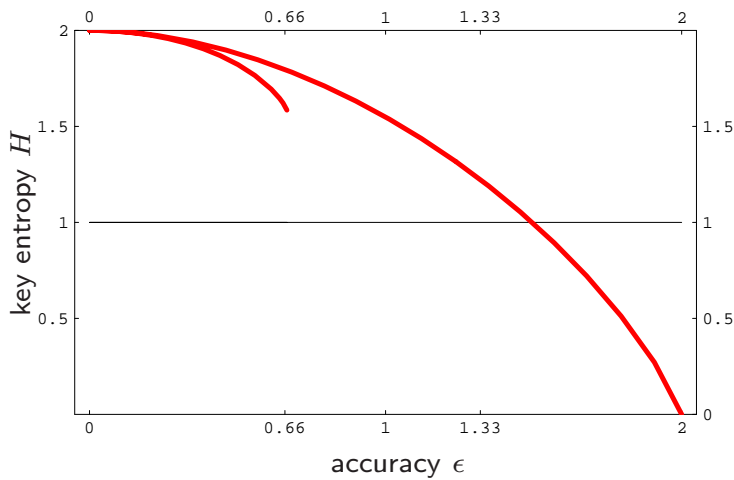}
\caption{The dependence of the entropy of the optimal key $H$ for
a given degree of secrecy $\epsilon$ is depicted for all depolarizing
channels characterized by $\lambda_x=\lambda_y=\lambda_z=\lambda\in[-1/3,1]$.
}
\end{center}
\end{figure}

While the depolarizing channels are optimal in the region of
channels with the positive $\lambda$s for each value of $\epsilon$,
in the negative region these channels are optimal only for
$\epsilon\in[0,2/3]$, because for other values the ``depolarizing''
channels are not physical. Let us remark that in the previous paragraph
we have discussed the optimality for depolarizing channels, but
for $\epsilon>2/3$ these are no longer optimal. We have found numerically
that the channels parametrized as follows
$\lambda_z=-\lambda$ (also for $\lambda\ge 1/3$) and
$\lambda_x=\lambda_y=-\kappa$ ($\kappa\le\lambda$) are optimal.
The inequalities (\ref{ineq}) results in the bound $\kappa\le (1-\lambda)/2$,
which is nontrivial only if $\lambda>1/3$. In fact, only in this case
the condition $\kappa\le\lambda$ holds. The minimal entropy for
$\lambda>1/3$ is achieved for $\kappa=(1-\lambda)/2$ when
\be
H_{II}=\frac{1}{2}\left[3+\lambda
-(1-\lambda)\log(1-\lambda)-(1+\lambda)\log(1+\lambda)
\right]
\ee
and $\epsilon=2\lambda$. For a given $\epsilon$ this function
should be compared with the entropy for depolarizing channel
$H_I$ for $\lambda\ge 1/3$. If $|\lambda|=0.4913$ ($\epsilon=0.9826$)
these two function coincides, i.e. $H_I=H_{II}$.

\begin{figure}
\begin{center}
\includegraphics[width=10cm]{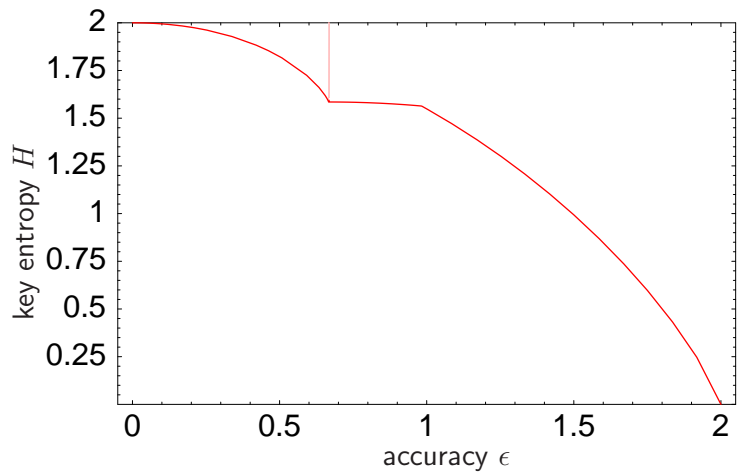}
\caption{The optimal entropy of the key $H$ with respect to
the accuracy $\epsilon$ for a single qubit 
aproximate private quantum channels. The optimal values are achieved
for depolarizing channels except the interval $2/3\le\epsilon\le 0.9826$, when
the optimal APQC is given by phase damping channels.}
\end{center}
\end{figure}

We have found that the optimal
entropy for a given value of the security parameter
is given by the following function (see Fig.3)
\be
H=\left\{
\ba{lcl}
H_1(-\epsilon) & {\rm for} & 0\le\epsilon\le 2/3 \\
H_2(\epsilon) & {\rm for} & 2/3\le\epsilon\le 0.9826 \\
H_1(\epsilon) & {\rm for} & 0.9826\le\epsilon\le 2 \\
\ea
\right. \, ,
\ee
where $H_1(\epsilon)=2-\frac{1}{4}[
(1+\frac{3}{2}\epsilon)\log(1-\frac{3}{2}\epsilon)
+3(1-\frac{\epsilon}{2})\log(1-\frac{\epsilon}{2})]$
and $H_2(\epsilon)=\frac{1}{2}[3+\frac{\epsilon}{2}-
(1-\frac{\epsilon}{2})\log(1-\frac{\epsilon}{2})-
(1+\frac{\epsilon}{2})\log(1+\frac{\epsilon}{2})]$.
This function characterizes optimal approximate
private quantum channels for the encryption of the whole state
space. To discuss the optimality of APQC for general sets
of plaintexts is beyond the scope
of this paper. The main obstacle is that the analysis cannot be reduced to
some typical sets like it was in the case of perfect PQC. Let us note
that for APQC not all the tracepreserving linear
combination must be encrypted with the given
security $\epsilon$. Similar result for approximatively private quantum
channels has been derived also in Ref.\cite{nagaj}, where also the optimality
for qubit was discussed using different methods and slightly different
definition of the security parameter $\epsilon$.

\section{Conclusion}
For single qubits we have characterized and analyzed
all possible ancilla-free private quantum channels and all possible approximate
private quantum channels.
We have shown that except the set of plaintexts $\cal P$ generating the whole
set of states via tracepreserving linear combinations, i.e.
if $\overline{\cal P}_{tp}\ne {\cal S}({\cal H})$, for arbitrary set
of plaintexts one bit of classical key is sufficient to establish
a private quantum channel. However,
if $\overline{\cal P}_{tp}={\cal S}({\cal H})$ then
two classical bits are necessary. In order to
use single bit of key even in such case, one should
employ an unphysical operation - universal NOT (${\cal E}_{\rm NOT}$).
The encryption of single qubit based on the operations
${\cal I},{\cal E}_{\rm NOT}$ (with equal probabilities)
would map arbitrary input state into the total mixture
$\varrho^{(0)}=\frac{1}{2}({\cal I}[\varrho]
+{\cal E}_{\rm NOT}[\varrho])=\frac{1}{2}(\varrho+\varrho^\perp)
=\frac{1}{2}I$.

Except the results valid for single qubit private quantum channels
we have derived bound on the optimal entropy for arbitrarily system.
In particular we have shown that for
PQC $[{\cal P},{\cal E},\varrho^{(0)}]$ the entropy of the key cannot
be smaller than the entropy exchange
$H\ge \max_{\varrho}S_{\rm ex}(\varrho,{\cal E})$. We have also shown that
for a given random unitary channel ${\cal E}$ the decomposition into mutually
orthogonal unitaries optimizes the entropy $H$. For qubits such decomposition
always exists, but for larger dimensional systems its existence
is an open problem. Except extending the results to larger systems,
it would be of interest to perform similar analysis for
private quantum channels involving the usage of ancilla in 
its full generality, i.e. including ancilla. This opens a lot of new
possibilities and some improval is very likely to happen.

\section*{Acknowledgement}
The work was supported by the project GA\v CR GA201/01/0413. MZ acknowledges
the support of Slovak Academy of Sciences via the project CE-PI, APVT-123 and
project INTAS (04-77-7289). J.B. acknowledges support of the Hertha Firnberg ARC stipend program and grant project GA\v CR 201/06/P338.


\section*{References}
\begin{thebibliography}{10}
\bibitem{gisin01}
N.Gisin, G.Ribordy, W.Tittel, and H.Zbinden,
Quantum Cryptography,
Rev.Mod.Phys., quant-ph/0101098

\bibitem{bouda04}
J.Bouda,
Encryption of quantum information and quantum cryptographic protocols,
PhD thesis, Faculty of Informatics, Masaryk university (Brno, 2004)

\bibitem{gottesman00}
D.Gottesman and H.K.Lo,
From quantum cheating to quantum security,
Physics Today 53, November (2000)

\bibitem{bennett84}
C.H.Bennett and G.Brassard,
Quantum cryptography: public key distribution and coin tossing,
Proceedings of IEEE Int. conference on computers, systems and signal
processing, Bangalore, India, pp. 175-179 (1984)

\bibitem{ekert91}
A.Ekert,
Quantum cryptography based on Bell's theorem,
Phys.Rev.Lett. 67, 661 (1991)

\bibitem{hillery99}
M.Hillery, V.Bu\v zek, and A. Berthiaume,
Quantum secret sharing,
Phys.Rev.A 59, 1829 (1999)

\bibitem{cleve99}
R.Cleve, D.Gottesman, and H.K.Lo,
How to share a quantum secret,
Phys.Rev.Lett. 85, 648-651 (1999)

\bibitem{bennett91}
C.H.Bennett, G.Brassard, C.Cr\'epeau, and M.H.Skubicziewska,
Practical quantum oblivious transfer,
Proceedings of the 11th Annual International Cryptology Conference on Advances in Cryptography, pp. 351-366 (1991)

\bibitem{crepeau94}
C.Cr\'epeau,
Quantum oblivious transfer,
J.Mod.Opt. 41, 2445-2454 (1994)

\bibitem{gruska99}
J.Gruska,
Quantum Computing,
(Osborne McGraw-Hill, 1999)

\bibitem{ambainis00}
A.Ambainis, M.Mosca, A.Tapp, and R.de Wolf,
Private quantum channels,
FOCS 2000, pp. 547-553 (2000)

\bibitem{boykin00}
P.O.Boykin and V.Roychowdhury,
Optimal encryption of quantum bits,
quant-ph/0003059

\bibitem{oppenheim03}
J.Oppenheim and M.Horodecki,
How to reuse a one-time pad and other notes on authentification, encryption and protection of quantum information,
quant-ph/0306161

\bibitem{leung00}
D.W.Leung,
Quantum Vernam cipher,
Quantum Information and Computation 2, 14-34 (2002)

\bibitem{barnum02}
H.Barnum, C.Cr\'epeau, D.Gottesman, A.Smith, and A.Tapp,
Authentification of quantum messages,
FOCS 2002 (2002), quant-ph/0205128

\bibitem{gottesman99}
D.Gottesman,
On the theory of quantum secret sharing,
Phys.Rev.A 61, 042311 (2000)

\bibitem{divincenzo02}
D.P.Di Vincenzo, P.Hayden, and B.M.Terhal,
Quantum data hiding,
Found.Phys. 33, 1629-1647 (2003)

\bibitem{werner01}
M.Gregoratti, R.F.Werner,
Quantum lost and found,
quant-ph/0209025

\bibitem{nayak}
A.Nayak, P.Sen,
Invertible quantum operations and perfect encryption of quantum states,
Quantum Information and Computation 7, No.1, 103-110 (2007)

\bibitem{perez}
A.Perez,
Quantum theory: concepts and methods,
(Kluwer, 1999)

\bibitem{ruskai}
M.B.Ruskai, S.Szarek, E.Werner,
An Analysis of Completely-Positive Trace-Preserving Maps on 2x2 Matrices,
Lin. Alg. Appl. 347, 159--187 (2002)

\bibitem{bouda05}
J.Bouda and M.Ziman,
Limits and restrictions of private quantum channel,
quant-ph/0506107 (2005)

\bibitem{ziman05}
M.Ziman, V.Bu\v zek,
All (qubit) decoherences: Complete characterization and physical implementation,
Phys.Rev.A 72, 022110 (2005)

\bibitem{nagaj}
I.Kerenidis, D.Nagaj,
On the optimality of quantum encryption schemes,
J. Math. Phys. 47, 092102 (2006)

\end{thebibliography}
\end{document}